# UNDERGROUND CURRENT IMPULSES AS A POSSIBLE SOURCE OF UNIPOLAR MAGNETIC PULSES

P. Nenovski

University Centre for Space Research and Technology, Sofia University, Bulgaria

**Abstract**

Recently several cases of observations of unipolar magnetic field pulses associated with earthquakes at different points (California, Italy, Peru) have been recorded. The paper attempts to model unipolar magnetic field pulses based on one mechanism that should be omnipresent for all measurement points, namely, the magnetic field diffusion through a conductive medium. The structure of magnetic fields supported by electric current sources is thoroughly modelled. The source of electric current is considered as an elongated volume of finite cross-section being immersed in a conductive medium. To model the unipolarity feature of the observed pulses prior to and at the earthquake main shock, the electric current of the source is of impulse form. Special attention is paid to the differences in the pulse structure (as amplitude envelope and the pulse width) that are measured by various magnetometers (fluxgate or search-coil). An analysis and comparison with recorded magnetic field pulse characteristics reveal that the observed unipolar pulses may have a common genesis, an electric current source within a conductive medium such as the earth crust.

# 1. Introduction

In the last two or three decades, a lot of papers have been devoted on identification of possible pre-earthquake anomalous geomagnetic field signals. Special efforts have been made to study electromagnetic signals/noise in the Ultra-Low-Frequency (ULF) band ranging from a few mHz to tens of Hz (see for example Hayakawa and Molchanov (2002), Molchanov and Hayakawa (2008), Hayakawa (2013), etc.).

Among the various aspects of the ULF signals associated with earthquakes, unipolar magnetic field pulses have been recently identified in magnetic field data of California, Italy, Peru, and other areas (Bleier et al., 2009; Villante et al, 2010; Bleier et al., 2012; Nenovski et al. 2013; Nenovski, 2015). These magnetic field signals represent relatively intense sharp pulses (either positive or negative) of specific shape and polarization. The pulses appear intermittently in a cluster form (Velante et al, 2010). Their pulse width is of several seconds and in the most cases, much less (Bleier et al, 2009; Villante et al, 2010; Dunson et al., 2011; Nenovski et al., 2013).

According to Dunson et al. (2011), three-axis induction (search-coil) magnetometers that are buried 1–2 m below the Earth's surface have recorded predominantly unipolar magnetic field pulses with durations of 0.1 to 12 s. They are stronger in the two horizontal (north-south and east-west) channels than in the vertical channels. The pulses last for many hours. They have large amplitudes and the amplitude ratios taken from pairs of orthogonal coils are stable across the bursts, suggesting a similar source.

A similar analysis of fluxgate and induction (search-coil) magnetometer data from the 2009 Aquila earthquake series was conducted by Villante et al. (2010). The researchers focussed on the daily rate of occurrence of negative or positive pulses of short duration. The typical duration of the detected pulses was shorter than 5 s. In the daily rate of that pulses, a general trend toward increasing in the course of time with approaching of the magnitude M6.2 earthquake was established (Villante et al, 2010). More precisely, maximum values in the D (east-west) component were observed about 20 days before the earthquake, on the 18$^{th}$ and 19$^{th}$ of March 2009. On 19$^{th}$ of March in substantial association with the lightning occurrence, sharp pulses were detected from both the induction and the fluxgate magnetometer. The pulses detected on March 18, in the absence of lightning activity, reveal different features: stable polarization and a tendency to be clustered in short time intervals of ~ 20 minutes length (Villante et al, 2010; Nenovski et al., 2013). One preliminary conclusion was stated by Villante et al (2010) that other (unknown) natural sources or artificial contamination may be the cause of their appearance. Note that the unique occurrence of unipolar magnetic pulses was checked over 3-year data before the Aquila earthquake (Villante et al, 2010).

In the literature there are additional findings that could shed light on the possible genesis of these unipolar pulses. For example, in the week prior to the Alum Rock M5.4 earthquake, Christman et al (2012) have revealed that their closest station (41 km from the epicentre) had recorded pulsations of similar duration and polarity to those identified by Bleier et al. (2009); the pulses did not appear to show any relationship with the forthcoming earthquake (e.g., more frequent or larger pulses, or a change in the type of the pulses, nearer in time to the earthquake) as has been reported by Bleier et al. (2009). Christman et al (2012) honestly mention "the difference in our findings from those of Bleier et al (2009) may result from the fact that the station was significantly further from the epicentre than theirs. Other factors may include that 1) we had significantly fewer pulse events to analyse, 2) we had studied a shorter window of time prior to the earthquake."

In order to explain the observed intense unipolar magnetic pulses (Bleier et al., 2009) as natural ones, Bortnik et al (2010) suggested that such magnetic signals are generated by a small current element co-located with the earthquake hypocentre radiating electromagnetic wave of arbitrary frequency $f$. The underlying mechanism of such current elements is not considered. For typical values of ground conductivity, the minimum electric current required to produce an observable magnetic field signal (of 1 Hz) within a 30 km range was found to be ~ 1 kA. Consequently, the authors underscored the importance of using a network of magnetometers when searching for magnetic field signals generated by underground electric currents. Unfortunately, the authors did not take into account that the unipolar magnetic pulses are essentially transient events and have nothing to do with continuous wave events, in other words, the proper analysis of unipolar pulses should be done in a time domain.

Fenoglio et al (1995) were the first to suggest that underground electric current sources of impulse type might be associated with earthquake processes and performed their analysis in a time domain. Similar approach has been applied by Surkov et al. (2003), Scoville et al. (2015a) and Yamazaki (2016). Surkov et al (2003) have elaborated the microcrack mechanism (Molchanov and Hayakawa, 1995) and have found that the sign of the components of the electromagnetic fields generated by all microcracks should be the same regardless of their space orientation. This coherence effect occurs, because the effective magnetic moments of the microcrack current systems are always directed opposite the geomagnetic field vector. Their model predicts magnetic and electric perturbations to be of the order of 1–10 nT and 1–10 µV/m, respectively, at a distance 10–50 km from the epicentre (Surkov et al., 2003). In advance, we clarify that the prescribed polarization does not fit the observed polarization characteristics of the unipolar magnetic pulses (Nenovski et al, 2013; Nenovski, 2015).

Scoville et al (2015) have attempted to model unipolar pulses assuming positive (p-) hole (electronic charge) generation mechanism in the Earth's crust (Freund, 2002). The magnetic field signals then appear as a consequence of the stress-induced release, diffusion and relaxation of unbound p-holes through the boundary between the stressed (source) and the surrounding unstressed (receiver) volumes. The suggested diffusion mechanism thus is of internal origin, realized within the rock volumes undergoing stress changes. Their model attempts to explain in a qualitative way the observed unipolar magnetic field pulses.

Another study worthy of being mentioned is the study of Yamazaki (2016). In it, the author develops analytical expressions for the magnetic field generated by an impulse line-current immersed in a conductive medium: whole-space, half-space or two-layer models. He concludes that the expected amplitudes of the electromagnetic signal would be within the detection limits of the commonly used sensors under the condition that ground conductivity is not very high and the source current is sufficiently intense.

In all previous analyses, the electric current impulses have been estimated applying the line current approach, i.e. the corresponding studies do not take into account the real geometry (volume and cross section) of the possible electric current sources. This circumstance may become important in the cases of measurements conducted in close vicinity to the epicentres of strong earthquakes. Rock volumes (possibly involved in the earthquake preparation processes) are considered as a source and their sizes may happen comparable to the distance between the source centre and the measurement point.

**2. Unipolar pulses and their characteristics**

Let us first summarize the main characteristics of the observed unipolar magnetic field pulses. Figure 1 (panels 1-3 and 5-7) illustrates the unipolar pulse activity recorded by fluxgate magnetometers on March 18, 2009 (Nenovski et al, 2013). The magnetic field pulses are mainly of dichotomous type or spikes. The pulse activity appears as clusters of several pulses of different width (1 to several seconds) and amplitude. The total amplitude reaches 1 nT. Figure 1, panel 4 includes also concomitant data of 1 s overhauser magnetometer. One may see simultaneous appearance of unipolar pulses in the fluxgate and the overhauser magnetometer (marked by ellipses). At 14:32 UT, there are two consecutive transient signals, the first one has peak amplitude of 1.2 nT, the second one has somewhat less amplitude (0.7 nT). Their width is within the range from 1 to several seconds. Another sequence of transient signals can be suspected by hints of pulses (denoted by another ellipse). They look like teeth.

The next figure (Figure 2) represents the transient signal observed by the overhauser magnetometer sketched in an exaggerated form (Fig. 1, panel 4). Pulses of step-like form may be seen around the base line (F = 46378.2 nT, Fig. 2). These unipolar signals create the impression that they are of transient nature. Their amplitude shape is similar to that examined by Nenovski (2015) (Figure 3). So far, the unipolar pulses recorded around the 2009 Aquila earthquake (Italy) are the only supported by observations with different types of magnetometers.

Unipolar pulses are first observed in California and Peru using only search-coil magnetometers (Bleier et al., 2009, 2012). Specifically, these pulses have a shorter width. Another specific difference of them compared to magnetic field pulses observed in Italy is that their amplitude shape is not exactly unipolar (Fig. 4). Its envelope consists of a sharp peak immediately followed by an "indention" of reverse sign of much less amplitude.

In overall, unipolar pulses measured by various magnetometers look as pulses with a steep rise usually within 1 s, a peak and slower decrease in amplitude. The average width of the unipolar pulses measured by search-coil magnetometers is within 0.2 s and after their peak (either positive or negative) the pulses evolve in indention of reverse sign and considerably reduced amplitude (Fig. 4). The recordings of unipolar pulses in California and Peru however are not supported by other types of magnetometers. To emphasize, the unipolar signals being recorded by fluxgate (and overhauser) magnetometers (Villante et al, 2010; Nenovski et al. 2013) represent a truly unipolar structure with wider pulse width (up to a few seconds) in comparison to pulses recorded by search-coil magnetometers.

Table 1 summarizes the unipolar pulse activity observed around the 2009 Aquila earthquake. It includes records of unipolar pulses around the 2009 Aquila earthquake area measured by overhauser and fluxgate magnetometers. The table presents supporting seismic information about processes concerning the time delay of arrival of S waves with different polarization presumably due to crack formation process under stress. Details are given in paper by Nenovski et al. (2013).

To determine whether the recorded unipolar magnetic pulses are natural, or artificial, an in-depth study of the excitation of underground electric currents, as well as knowledge of their spatial and temporal characteristics at the Earth's surface are needed. To evaluate the importance of underground electric currents of impulse type originating from potential sources as nucleation zones of impending earthquakes, as well their possible effects, one needs a realistic theoretical model of such currents. This means a detailed study of their magnetic pulse characteristics reflecting all physical parameters (as finite geometry, realistic charge densities and velocities) of the impulse source and their dependence on crust parameters, i.e. the electromagnetic properties of their surroundings. Knowing the possible characteristics of underground currents, special prescriptions to the measurement methodology and techniques (including proper sampling frequency, pulse profiles, etc.) for searching seismic-related unipolar pulses in the ULF range may be made. It is also possible to identify favourable locations for monitoring of such unipolar pulses.

The model of impulse current sources examined below (the next section), reveals unexpected so far, inherent properties, such as changeable pulse width that depends on the actual geometry of the current volume and its cross-section. It turns out that the pulse width differs depending also on the measuring technique and hence, unipolar magnetic pulses will be recorded in different frequency ranges for different magnetometers. Finally, the maximum amplitude of the signal produced by impulse current source exceeds several times the amplitude of the current sources of invariable strength with time. A comparison of the model with available data of unipolar pulses reveals that the observed magnetic field pulse series may be associated with sources of impulse currents immersed in a conductive medium of conductivity values close to that of the earth crust.

**3. Model of transient electric current source.**

The experimental evidence of unipolar pulses (from different continents) suggests a common mechanism: an impulse electric current of finite size through a conductive medium and subsequently diffused through the Earth's crust.

We model an electric current generated within a volume of finite geometry and its magnetic field irrespective of the generation mechanism. For that purpose, an impulse electric current is assumed to be triggered in some volume $V$ at some moment $t$. This volume is surrounded by a medium of electrical conductivity $\sigma$. The electric current produces its own magnetic field which is subject to a diffusion process through this medium. The magnetic diffusion time is characterized by the factor $r^2\mu\sigma$ where $r$ is the distance, and $\mu$ is the magnetic permeability (e.g. Shkarofsky et al, 1966; Boyd and Sanderson, 1969). The time scale of the electric current impulse itself depends on the type of the mechanism. For this purpose, the impulse current duration $\tau_i$ is assumed to be much less than the expected diffusion time of the medium, say $t_o \approx r^2\mu_0\sigma$. Assuming $\tau_i \ll \tau_o$, the impulse current may be easily represented by the Dirac delta (impulse) function $\delta(t)$. Further, the source electric current density $j_0$ is assumed to occupy a channel instead of dimensionless lines. The assumed channel has a cross section of rectangular geometry with dimensions $2x_0 \times 2y_0$, where $x_0$ and $y_0$ may vary within a wide scale (Fig. 5). The assumed channel is oriented along the $z$ axis and, for convenience, is considered to be of infinite length. The amplitude of the electric current is assumed constant along the $z$ axis, which is reflected by the formula

$$j_0(x,y) = j_0(\theta(x+x_0) - \theta(x-x_0))(\theta(y+y_0) - \theta(y-y_0))\delta(t) \qquad (1)$$

where $\theta(x)$ is the Heaviside step function; and $2x_0$ and $2y_0$ are the transverse scales of the current source (Fig. 5, Table 2).

Further, the source electric current density is represented by $\vec{j} = j_0\hat{z}$, where $j_0$ is the current amplitude and $\hat{z}$ is the unit vector along the current direction. The electric current thus produces a transient magnetic field in the surrounding medium of conductivity $\sigma$. The initial equation governing the magnetic field behaviour in time and space is:

$$-\Delta\vec{B} + \mu_0\sigma\frac{\partial\vec{B}}{\partial t} = \mu_0 \, grad \, j_0 \times \hat{z} \qquad (2)$$

where $B$ is the magnitude of the magnetic field produced in the surrounding medium by the electric current of density $j_0$.

This equation directly follows from Maxwell equations assuming that the displacement current, defined as $\varepsilon\frac{\partial E}{\partial t}$, is much smaller compared to the conductivity current density, $\vec{j}_{con}$ defined as $\vec{j}_{con} = \sigma\vec{E}$, where $E$ is the electric field generated in the medium surrounding the volume of the driver current. This means that the time variations of the electromagnetic field are assumed to be slow and thus, the inequality $\vec{j}_{con} \gg \varepsilon\frac{\partial E}{\partial t}$ is well satisfied. Having in mind the time scales observed at earthquake shock and before it, the above inequality holds for the given conductivity of the medium. Note that on the right-hand side of the equation there is only the driver part of the electric current density $\vec{j}$. The conductivity part $\vec{j}_{con}$, is incorporated implicitly into the left-hand side of equation (2).

The governing equation (2) needs to be rewritten in non-dimensionless variables. Using $r$ where $r$ refers to the distance between the source and the point of measurement and assuming $\tau_0 \equiv \mu_0\sigma S$ ($\mu_0$ is the magnetic permeability of vacuum, $S \equiv r^2$) as a regulatory factor, the governing equation can be rewritten as:

$$\left(-\frac{\partial^2}{\partial \tilde{x}^2} - \frac{\partial^2}{\partial \tilde{y}^2} + \frac{\partial}{\partial \tilde{t}}\right) B_x = \mu_0 j_0 \sqrt{S} (\delta\left(\tilde{y} + \frac{y_0}{x_0}\right) - \delta\left(\tilde{y} - \frac{y_0}{x_0}\right))(\theta(x_0(\tilde{x}+1)) - \theta(x_0(\tilde{x}-1)))$$

$$\delta(\frac{t}{\tau_0})$$

(3)

where $\tilde{x} = \frac{x}{\sqrt{S}}$, $\tilde{y} = \frac{y}{\sqrt{S}}$, $\tilde{t} = \frac{t}{\tau_0}$ and

$$\left(-\frac{\partial^2}{\partial \tilde{x}^2} - \frac{\partial^2}{\partial \tilde{y}^2} + \frac{\partial}{\partial \tilde{t}}\right) B_y$$

$$= -\mu_0 j_0 \sqrt{S} (\delta(\tilde{x}+1) - \delta(\tilde{x}-1))(\theta\left(x_0\left(\tilde{y} + \frac{y_0}{x_0}\right)\right)$$

$$- \theta\left(x_0\left(\tilde{y} - \frac{y_0}{x_0}\right)\right))\delta(\frac{t}{\tau_0})$$

(4)

The solution is sought applying the Green function approach (e.g. Bayin, 2006). The following solution of the governing equations is thus derived:

$$B_x = \frac{1}{\sqrt{\pi}} \left(\frac{\mu_0 j_0 r}{\sqrt{\tilde{t}}}\right) sh\left(\frac{\tilde{y}\tilde{y}_0}{2\tilde{t}}\right) \exp\left(-\frac{\tilde{y}^2}{4\tilde{t}}\right) \left(\text{erf}\left(\frac{\tilde{x}+\tilde{x}_0}{2\sqrt{\tilde{t}}}\right) - \text{erf}\left(\frac{\tilde{x}-\tilde{x}_0}{2\sqrt{\tilde{t}}}\right)\right) \quad (5a)$$

and

$$B_y = \frac{1}{\sqrt{\pi}} \left(\frac{\mu_0 j_0 r}{\sqrt{\tilde{t}}}\right) sh\left(\frac{\tilde{x}\tilde{x}_0}{2\tilde{t}}\right) \exp\left(-\frac{\tilde{x}^2}{4\tilde{t}}\right) \left(\text{erf}\left(\frac{\tilde{y}+\tilde{y}_0}{2\sqrt{\tilde{t}}}\right) - \text{erf}\left(\frac{\tilde{y}-\tilde{y}_0}{2\sqrt{\tilde{t}}}\right)\right) \quad (5b)$$

where $\tilde{x}$ and $\tilde{y}$ are the dimensionless coordinates of the point $r$ ($r \equiv (x,y)$) where the magnetic field $\vec{B}$ is measured; $\tilde{x}_0$, $\tilde{y}_0$ are the dimensionless cross-section half sides of the current channel dimensions ($x_0$, $y_0$). Actually, they are equal to 1 but retained in eqs. (5a) and (5b) for convenience. Expressions (5a) and (5b) yield the magnetic field at any point ($x,y$) produced by a current channel of finite cross section (2$x_0$, 2$y_0$).

Table 2 summarizes all the spatial and temporal scales involved in the model.

The suggested diffusion model simulates the magnetic field produced by a source of impulse electric current of finite cross-section but infinite length. As it was pointed out, the impulse current is initiated at some initial moment and disappears, being simulated by a delta function source. The amplitude profile of the magnetic field signal and its characteristics are different and specific, as well, depending on the source geometry (see formulae (A2a) and (A2b) in Appendix A. The next step is to study the effects of the finite cross-section source of impulse current on the pulse width and amplitude.

*3.1. Pulse width*

We simulate either a strip, or a squared cross-section of the impulse current source volume. Note that the strip current geometry model admits to varying the ratio $x_0/y_0$. The squared cross-section means that the cross-section sides of the source volume are equal, i.e. $x_0 \approx y_0$; In respective calculations, the squared cross-section of side $x_0$ is assumed (for convenience) equal to $0.01r$, where $r$ is the distance between the current source and the point of measurement. Figures 6 and 7 illustrate the pulse envelope differences between the strip and squared cross-section impulse current models (see (A3) and (A4) in Appendix A). The squared cross-section current reveals a lesser pulse width compared to that of the strip impulse current model. Note that the apparent width $w_m$ of the magnetic field is $0.268t_0$ (for impulse line current) or $0.45t_0$ (for impulse strip current). The pulse amplitude and their widths are expressed in normalized time coordinate, $t/t_o$ (Table 2). For example, at distance of 10 km, and a medium conductivity $\sigma$ of $10^{-2}$ S/m the diffusion time $t_o$ becomes close to 1 s.

To simulate signals measured by search-coil magnetometers, the pulses of magnetic field and magnetic field $B$ and its derivative ($dB/dt$) are plotted simultaneously (Fig 7 and Fig. 8). It is worth noting that in the low-frequency limit, search-coil magnetometers are directly measuring the magnetic field derivative ($\frac{dB}{dt}$) (see Appendix B). Thus, the apparent width $w_{md}$ of the magnetic field derivative $dB/dt$ amounts to $0.051t_0$ ($0.065t_0$) for impulse line (strip) impulse currents, respectively.

Figure 8 illustrates quantitatively the peak amplitude difference between the line current and squared cross-section currents calculated at their peak amplitudes. Cases of three diffusion times 0.05 s, 0.1 s and 0.5 s are considered. It can be noted that for impulse current sources of size less than 1 % of the distance $r$ between the source and the measuring point, the coincidence of the finite cross-section current model with the impulse line current is good. The dash line (Fig. 8) marks the limit of the line current concept.

The magnetic field pulse width will differ depending on i) the diffusion time, $t_0$, ii) the model geometry (squared or strip) and iii) the methodology of measurements.

*3.2. Pulse amplitude*

The pulse amplitude measured on the ground is crucial for the possible detection of current impulses generated within the earth's crust. In order to make some estimates of the expected magnitudes of the magnetic field signal, we assume again a reasonable distance $r$ of 10 km ($10^4$ m). Due to diffusion, the magnetic field amplitude reaches its maximum values at a time approaching the time $t_{peak} = \tau_d/8$ (squared cross section), or $\tau_d/6$ (strip). Note also an important finding that the peak amplitude of the diffusion signal is ~ 4.5 higher than the amplitude at the static current limit (Appendix A). Thus, at distances of $10^4$ m, the primary magnetic field $B_{max} = \sqrt{B_x^2 + B_y^2}$ (eqs (A3a) and (A3b)) may become as high as 1 nT, provided the current strength, $I_0$ ($I_0 \equiv 4j_0 x_0 y_0$), is ~15 A. In the static limit, greater current strength $I_0$ is necessary to produce 1 nT signal at a distance $r$ of $10^4$ m: ~ 65–70 A. Of course, this peak amplitude is reached under optimal conditions: when the measurement occurs at a plane perpendicular to the current direction and is centred at its volume.

It is worth noting that any current signal generated within the crust will be subject to reflection and transmission at the air-ground interface. The reflection and transmission coefficients depend on the crust conductivity (if air is assumed as an ideal dielectric). This problem has been treated by Wait (1982) assuming a line current buried in the crust. Following Wait's theoretical results for the reflection and transmission coefficients at the interface between semi-infinite media of uniform conductivities, the primary magnetic field reduces appreciably at the ground, while the corresponding electric field may increase up to twice. Transient magnetic signals however can transmit even high conductive layers (e.g. wet soils) without amplitude losses providing they are sufficiently thin compared to the thick high resistivity (air and rock) surroundings.

## 4. Analysis

So far, the unipolar pulses have been revealed using three different magnetometers: induction (search-coil) of high resolution (Bleier et al. 2009), fluxgate and overhauser magnetometers of 1 Hz sampling frequency (Villante et al, 2010, Nenovski et al, 2013). Their pulse widths were in most cases about 1 seconds (recorded by fluxgate and overhauser magnetometers) and much less than 1 second (recorded by search-coil magnetometers). Next step of analysis is to compare the suggested model to available data set of observed unipolar pulse series. i.e. If the unipolar magnetic field pulses are due to diffusion effect produced through the conductivity medium, then the apparent width values $w_m$ and $w_{dm}$ would characterize the distance or conductivity that the signal penetrate. Knowing the conductivity profile, one may evaluate the distance to the source and possibly its geometry. The fitting procedure will be done by the impulse finite cross-section (squared or strip) current model. A preliminary fitting procedure thus is applied on a pulse of available pulse series: that of greatest apparent pulse width (Fig. 2). Fortunately, for that pulse its amplitude profile consists of sufficient measuring points (over 50). The fitting procedure was done using the squared cross-section impulse current model. Figure 9 illustrates the experimental data (dotted line) and the squared cross-section model (solid line) that fits the data. The offset of the base line was taken into account. Note the observational points resemble a diffusion process with an apparent pulse width $w_m$ equal appr. to 5 s. Interestingly, the experimental data set of the magnetic field difference ($\Delta B$) is plotted (Fig. 10) and reveal an amplitude profile similar to that in Fig. 4 (gained by search-coil magnetometers). Note that there is a concavity of the experimentally derived magnetic field difference profile (dotted line) with a value of 0.2 of the normalized amplitude. The magnetic field derivative $dB/dt$ calculated from the squared cross-section current model fits in good consistency the magnetic field difference data set. Both fits are in agreement to each other. They infer a diffusion time in the range 10–20 s for that pulse. Unfortunately, more in-depth analysis of the whole unipolar pulse series was impossible. The reason is that the most of pulse series are of shorter widths (pulses consist of fewer points) and the fitting procedure (due to insufficient measuring points) was not applicable to that data sets.

According to the impulse current model, the arrival time of the pulse peak $t_{peak}$ is closely related to the diffusion time $t_0$. To simplify the task, let us apply the limit case – a line (l) or an impulse strip (s) impulse current (see Appendix A). The arrival time of the pulse peak $t_{peak}$ is related to the diffusion time $t_0$ by:

$$t_{peak,l} = t_0/8, \text{ or } t_{peak,s} = t_0/6 \tag{6}$$

i.e. the diffusion time $t_0$ is either 8, or 6 times the arrival time of the pulse peak $t_{peak,l(s)}$, respectively, depending on the impulse current geometry (line (l) or sheet (s)). There are experimental evidences of the possible arrival times of the recorded unipolar pulses. Due to the sampling frequency of 1Hz, the rise time of recorded (multiple) pulses, however is not recorded exactly although. This is within 1 s. Given that, it is assumed that the (maximum) arrival time $t_{peak}$ of the pulse peak is 1 s and less. The estimations of the respective diffusion time $t_0$ and parameters (distance r and conductivity σ) participating in $t_0$ are straightforward. Adopting the impulse current (squared cross-section or strip) geometry, we obtain the following relation:

$$\frac{t_0}{\mu_0} \equiv r^2\sigma \leq \frac{6(8)t_{max,s}}{\mu_0} \approx 5(6.5) \times 10^6 \tag{7}$$

The distance $r$ (entering in (7)) is tentatively assumed equal to $10^4$ m. Referring to both the 2007 Alum Rock and the 2009 Aquila earthquakes, the earthquake hypocentres are ins consistence with that distance. The conductivity $\sigma$ then (from (7)) would amount to $5 \times 10^{-2}$ S/m. Having in mind that the Aquila magnetic observatory was at distance of 6.7 km from the epicentre, the actual distance $r$ (to the earthquake nucleation zone) may be up to 12 km (1.2 $x10^4$ m). Under these assumptions, the estimated conductivity $\sigma$ would be even less, around $3.5 \times 10^{-2}$ S/m. Due to uncertainty in the arrival time $t_{peak}$ (< 1 s) the estimated conductivity value may be even less than $3.5 \times 10^{-2}$ S/m.

Generally, the crust conductivity is not uniform. As for the 2009 Aquila earthquake, there is an available 1-D model of the conductivity profile beneath the Aquila magnetic observatory. A multi-layered model has been accepted after respective telluric measurements (Di Lorenzo et al, 2011): 2 km upper layer of 5 Ωm, 3 km of 3000 Ωm, and 5 km of 500 Ωm in the first 10 km depth, etc. The earthquake hypocentre is close to the bottom side of the third layer, i.e. at a depth of 10 km. Provisionally, the total peak time $t_{peak}$ should be the sum of three peak times, i.e.:

$$r^2\sigma_{aver} \approx \frac{t_1+t_2+t_3}{\mu_0} \approx r_1^2\sigma_1 + r_2^2\sigma_2 + r_3^2\sigma_3 \tag{8}$$

from where we may calculate the average conductivity $\sigma_{aver}$ ~ 0.015 S/m. This conductivity estimate (based on the 1-D conductivity profile model) is less than 0.035 S/m, a value that is obtained based on the assumed uniform conductivity $\sigma$. Such a coincidence may be an indirect argument in support of a subsurface (underground) source of the observed magnetic pulses. The compliance thus can be assumed to exist. This assertion however, cannot be proven due to various reasons: i) unknown and unavailable distance $r$ between the measurement point and the source of impulse current, unknown and probably inhomogeneous conductivity distribution along the pulse propagation route to the measuring point (on the earth surface) and model simplifications, iii) the actual profile of the initial impulse current at the source, i.e. whether the pulse width is due to its own, internal diffusion process or due to the diffusion effect in the conductive medium through which the signal passes, etc.

Supposing we refer to search-coil magnetometer data set around the 2010 Tacna earthquake (Peru) (Dunson et al, 2011), the average pulse width is well determined (Fig. 4). Roughly the apparent pulse width $w_{md}$ is around 0.1 s. To remind that the search-coil magnetometers are measuring the magnetic field derivative (Appendix B). Following the modelled relationships between the apparent width $w_{md}$ and the diffusion time $t_0$ ($w_{md} = 0.051t_0$ or $0.065t_0$), the estimated diffusion time $t_0$ should be 1.5–2 s. Assuming that the pulse source is somehow associated with the earthquake nucleation zone, i.e. distance $r$ is presummed, we could estimate the average conductivity for that site. Following Dunson et al, (2011), the distance between the magnetic field measurement point and the 2010 Tacna earthquake epicentre was 37 km, the hypocentre depth – 30 km. Hence, the total distance $r$ is ~50 km. Using a diffusion time value of 2 s, the estimated conductivity of the upper part (up to 30 km depth) of the earth crust beneath Tacna should be close to $10^{-3}$ S/m. If we knew the conductivity profile (a preferred situation), we could estimate the distance $r$ (a requested parameter) to the source.

As for the observed co-seismic magnetic field pulse at the 2009 Aquila earthquake shock, the observed apparent pulse width $w_m$ is appr. of one minute (60−70 s, Nenovski, 2015). Applying the same approach (considering an appropriate strip impulse current model), the diffusion time $t_0$ may be calculated from the relation $w_m = 0.45t_0$ that yields a value of $1.5 \times 10^2$ s. Hence, the estimated conductivity would be ~ 1.5 S/m. That conductivity value is impossible for the Aquila observatory site placed on mountain basin. The observed co-seismic diffusion process, hence, should be explained by different mechanism, e.g. some (still unknown) diffusion process within the pulse source itself. Some of them are electrokinetic (e.g. Gershenzon et al. 2014), frictional (Leeman et al, 2014), etc.

It is worth noting here that the reliability of the observed co-seismic signal was disputed by Masci and Thomas (2016). Their inference for an artificial origin of the observed co-seismic signal are not based on a well-grounded analysis of all accessible magnetic data. The authors have previously cut off a 37 second interval in which is actually found the amplitude peak of the observed co-seismic pulse and its body.

The above analysis and preliminary comparison with unipolar pulse series recorded around the 2009 Aquila earthquake reveal that the magnetic diffusion through conductive medium may be a reliable mechanism that modifies the magnetic field of impulse current excited in the same medium. The results of the analysis of the structure of the magnetic field derivative also suggest that unipolar pulses recorded by search-coil magnetometers (in California, Peru, etc) may have the same origin, i.e. impulse currents generated within the conductive earth crust.

There is another aspect that may have a direct bearing on the proposed model. Previous estimates of the amount of electrical charges and currents produced with each of the mechanisms known so far have not been done on basis of the actual volume and geometry of their sources in the earth's crust. With the proposed model of the magnetic field structure from impulse current sources with real geometry, the updated estimates of the amount of possible charges and currents would be able to re-start on real (actual) cross-section sizes and volumes. By modeling the shape of the amplitudes and the pulse width one may derive additional information about the volume and the sizes of the pulse sources. This approach is similar to conventional geophysical prospecting. This would bring us closer to solving the most important task in this topic: the respective mechanisms can be quantified and tested for their availability.

Once unipolar magnetic field pulses have been allowed to be associated with sources in the earth's crust, it remains to discuss their possible mechanism. Whether is possible really to generate current impulses of extremely short duration, e.g. $10^{-4}$–$10^{-1}$ s? One answer to this key question is that in rocks, various mechanisms of nearly abrupt charge separation are possible, for example, such as (micro)crack formation processes (see for instance, Enomoto and Chaudhri, 1993; Fifolt et al, 1993, Gokhberg et al. 1985, etc). Electric charges can be released (freed) accompanying the crack tip propagation velocity being of about 1 km/s and separated (effectively) from bound charges of opposite sign. A proper consideration of such mechanisms and associated current source and geometry however requires another study that presently is not included.

Presumably, freed electric charges (as carriers of an impulse electric current under appropriate conditions) are released owing to external forces (e.g. stress changes). Expectedly, this process arises deep in the Earth's interior and possibly close to the associated earthquake nucleation zone.

Given the short pulse widths of pulses possibly generated by impulse currents in a conductive medium, it is recommended the following prescription for searching seismic-related pulse signals. At first, the sampling frequency of magnetic field measurements should be much higher than is now running. The conventionally perceived sampling frequency of 1 Hz is quite insufficient, especially if the measurements are conducted with (preferred) search-coil magnetometers. Presently, the magnetic field measurements conducted by (Bleier and his Quakefinder team) are only adequate in searching seismic-related signals.

## 5. Conclusion.

The modelling of impulse current of various geometry done in the present paper attempts to demonstrate that unipolar magnetic field pulses of short duration may come from a source within the earth crust. An electric current flowing through an elongated volume of finite cross-section being immersed in a conductive medium is modelled. Its magnetic field magnitude and pulse structure are analytically and numerically derived and evaluated. The assumed electric current is of impulse form. The finite cross-section geometry of the current sources modifies the pulse characteristics. Special attention is paid to changes of the pulse amplitude envelope and width that are measured by various magnetometers (fluxgate or search-coil). The modelled magnetic fields amplitude and shapes depending on cross-section geometry generalize the impulse line current concept. The geometry of the source of impulse current can be of importance, especially if strong earthquakes are considered because the measuring points easily may fall within the preparation zone affected by that earthquake. A preliminary analysis and comparison with accessible recorded magnetic field pulse characteristics reveals that the observed unipolar pulses may have really a common genesis within conductive medium such as the earth crust.

Given the rarely observational evidences of unipolar magnetic pulses presented so far, the true genesis of the impulse currents as a source of unipolar pulses remains still unknown. Experimental data, geology and the proposed modelling together would allow to estimate precisely the required carrier charge and current densities within the earthquake nucleation volumes needed to generate magnetic pulses of measurable magnitude and shape. Then, previously proposed mechanisms may be considered to test, verify and validate their compliance with the recorded unipolar pulse characteristics. A further examination of proposed mechanisms and generated electric charges and electric currents as a source of the observed unipolar pulses will be addressed in detail in a subsequent paper.

**Acknowledgement** The author is grateful to professors S. Shanov and F. Freund, Dr T. Bleier and Dr R. Glavcheva for stimulating discussions on the topic, their comments and invariable support.

# APPENDIX A

The distance *r* between the measurement point *r(x,y)* and the axis of the source electric current (placed at point (*x*=0,*y*=0)) may exceed the *cross section* parameters $x_0$ and $y_0$ ($r \gg x_0$ or $y_0$). If $x_0$ (or $y_0$) $\gg y_0$ (or $x_0$), a strip current geometry (*fault-like plane*) is formed. Otherwise, if the cross section parameters are of comparable size, i.e. $x_0 \approx y_0$, and the distance *r(x,y)* exceeds $x_0, y_0$, the error function in expressions (5) may be safely approximated by:

$$\mathrm{erf}(u \pm \delta u) \approx \mathrm{erf}(u) \pm \exp(-u^2)\delta u \tag{A1}$$

where *u* is the function argument: *u* is equal to $\frac{\tilde{x}}{2\sqrt{\tilde{t}}}$ or $\frac{\tilde{y}}{2\sqrt{\tilde{t}}}$; $\delta u$ is equal to $\frac{\tilde{x}_0}{\sqrt{\pi \tilde{t}}}$ or $\frac{\tilde{y}_0}{\sqrt{\pi \tilde{t}}}$, respectively. The final expressions are:

$$B_x = \frac{2}{\pi}\frac{\mu_0^2 \sigma j_0 x_0 R^2}{t} sh\left(\frac{\tilde{y}\, y_0}{2\tilde{t}\sqrt{S}}\right)\exp\left(-\frac{\tilde{x}^2+\tilde{y}^2}{4\tilde{t}}\right) \tag{A2a}$$

$$B_y = \frac{2}{\pi}\frac{\mu_0^2 \sigma j_0 y_0 R^2}{t} sh\left(\frac{\tilde{x}\, x_0}{2\tilde{t}\sqrt{S}}\right)\exp\left(-\frac{\tilde{x}^2+\tilde{y}^2}{4\tilde{t}}\right) \tag{A2b}$$

Let us introduce a diffusion time scale of medium conductivity σ: $\tau_d \equiv \mu_0 \sigma r^2$, where $r \equiv \sqrt{(x^2+y^2)}$. Let us denote $j_0(2x_0)(2y_0)$ as $I_0$ (the current source strength). Further, $I_0$ is considered *constant* irrespectively of the magnitudes of the actual cross-section parameters $(x_0, y_0)$. Then the magnetic field component $B_x$ varies as:

$$B_x = \frac{1}{2\pi}\left(\frac{\mu_0 I_0 \tau}{t y_0}\right) sh\left(\frac{\tilde{y}\, y_0}{2\tilde{t}r}\right)\exp\left(-\frac{\tilde{r}^2}{4\tilde{t}}\right) \tag{A3}$$

The magnetic field expression (when $x_0/y_0 \approx 1$) may be applied to a *square* or *cylindrical (tube)* current geometry of radius $x_o \approx y_o$. In that case, the azimuthal magnetic field component $B_\varphi$, supported by a current of cylindrical cross section of radius *R* will be expressed by (A3).

Let us now consider a *strip* current of width $2y_o$ and infinite length along the *x* axis ($x_0 \to \infty$). The electric current density $\vec{j}_0$ is parallel to the *z*-axis. Then:

$$B_x(y) = \frac{2}{\sqrt{\pi}}\left(\frac{\mu_0 j_0 y}{\sqrt{\tilde{t}}}\right) sh\left(\frac{\tilde{y}\, y_0}{2\tilde{t}y}\right)\exp\left(-\frac{\tilde{y}^2}{4\tilde{t}}\right) \tag{A4}$$

where *y* is the distance to the middle plane (defined as *y* = 0) of the strip current. This expression yields the magnetic field produced by electric current density $j_0$ along the *z* axis permeating an infinite strip current of thickness $2y_0$.

The strip current model may have various applications in geo-electromagnetism. One of them is electric currents that are released in soil through the earthing systems (this occurs mainly in a shallow layer of some depth of higher conductivity). Other strip (or layer) structures of high conductivity ( $\sigma \approx 10^{-1} \div 4$ S/m) are the fault, water basin systems (river, lake, sea, ocean), where electric currents (induced mainly by global geomagnetic activity) may be concentrated.

Expressions (3a and 3b) describe the magnetic field variations in space and time produced by transient electric currents of finite extension (along *x* and *y*). Naturally, these general expressions should be consistent with the expression of the magnetic field produced by a line current impulse. These expressions should also represent a generalization of the well-known formula of the magnetic field produced by a line current in static approximation ($t \to \infty$).

First, let us check the validity of (A3) at the limit ($y_0 \to 0$ and $x_0 \to 0$). Applying the L'Hôspital rule to expression (A2a), from straightforward algebra one gets:

$$B_{line,tr} = \frac{1}{4\pi} \frac{\mu_0 I_0}{r} \frac{1}{\tilde{t}^2} \exp(-\frac{\tilde{r}^2}{4\tilde{t}}) \tag{A5}$$

where $B_x$ is now replaced by the symbol $B_{line,tr}$ and the coordinate *y* – by *r*. Further, when there is a constant current source, e.g. switched on at time *t* = 0, the current source ~ δ(*t*) in (1) should be replaced by the Heaviside's function θ(*t*). Then, the magnetic field at moment *t* will be simply derived by integration over $\tilde{t}$:

$$B_{line} = \frac{1}{2\pi} \frac{\mu_0 I_0}{r} \exp\left(-\frac{r^2 \mu_0 \sigma}{4t}\right) \tag{A6}$$

For $t \to \infty$, expression (A6) expectedly reduces to:

$$B_{line,st} \to \frac{\mu_0 I_0}{2\pi r} \tag{A7}$$

which is identical to the magnetic field that is circumferential to an infinite wire carrying a current of strength $I_o$.

Attention should be paid to the ratio $B_{line,tr}/B_{line,st}$ and its time profile. At its maximum (when $t = t_{peak} = \tau_d/8$), this ratio becomes equal to ~ 4.5. This suggests that at the given measurement point the transient magnetic signal emerges with greater amplitude (4.5 times) compared to that produced by same current strength under static conditions.

# APPENDIX B

In fact, unipolar pulses have been recorded by several kinds of magnetometers, among them fluxgate and search-coils magnetometers. Fluxgate magnetometers are designed to *measure* DC fields, while the *searchcoil* – AC fields. Concretely, the search-coil sensor is built on the induction principle. The induced voltage, say $\epsilon$, is proportional to the time derivative of the magnetic flux as follows from the Faraday's law, i.e.

$$\epsilon = -nS\frac{dB}{dt} \qquad (B1)$$

where *S* is the cross-section and n is the number of turns in the sensor coil. The coil sensor is characterized by self-inductance *L*. It also features resistance *R* and capacitance *C*. Generally speaking, the transfer function between the output (the measurable voltage, $V_{out}$) and the flux density (the magnetic field, *B*) of the induction sensor depends on both the resonance frequency $\omega_r$ (equal to 1/sqrt(*LC*)) and the *RC* constant. However, their effects should be negligible under low-frequency conditions: $\omega \ll \omega_r$ and $\omega \ll 1/RC$ *(fulfilled for frequencies below 1 Hz, the fluxgate sampling frequency)*. This condition is practically satisfied when the frequency is less than 10–100 Hz. Note that the unipolar pulses observed by fluxgate magnetometers (of sampling frequency of 1 Hz) are well below this upper limit.

Examining unipolar magnetic pulses recorded by fluxgate magnetometers under these low-frequency conditions, the validity of eq. (B1) is confirmed.

Table 1. **Unipolar magnetic pulses (The M6.1 L'Aquila earthquake, 06 April 2009)**

|  | On 14 and 16 Feb | On 18 Mar | Earthquake main shock (06. Apr) | Remarks |
|---|---|---|---|---|
| Amplitude | ~ 0.2 nT | ~ 1 nT | 0.8 nT |  |
| Amplitude shape | Spikes or dichotomous | Spikes or dichotomous | Single unipolar signal | Recorded by fluxgate and overhauser magnetometers |
| Pulse width | 1–10 s | 1–5 s | 60–70 s |  |
| Pulse activity (duration) | Several hours | clusters of 20 min (intermittently) | ~ 5 min |  |
| Polarization (in horizontal plane) | Stable, Northeast−Southwest | Stable, Northeast−Southwest | Stable, Northwest−Southeast | Recorded by fluxgate magnetometers |
| Number of days before the EQ main shock | 50, 48 | 18 | 0 (at the EQ shock) |  |

Table 2 **Time scales and other parameters of the electric current source**

| Parameters | Definition |
|---|---|
| $\tau_i$ | Intrinsic time scale of the electric current impulse itself; s |
| $t_0$ | Magnetic diffusion time characterized by the factor $r^2\mu_0\sigma$, where $r$ is the distance between the source and the point of measurement, $\mu_0$ – magnetic permeability and $\sigma$ - electrical conductivity of the medium; s <br> Ref. Shkarofsky et al (1966); Boyd and Sanderson (1969) |
| $j_0$ | Source electric current density, A/m$^2$ |
| $2x_0, 2y_0$ | Dimensions of the sides of a rectangle cross-section of the electric current channel, m |
| $\vec{J}_{con}$ | Conductivity current density, $\sigma\vec{E}$; $\vec{E}$ is the electric field induced by the magnetic field variations $\vec{B}$; A/m$^2$ |
| $\tau_0$ | A regulatory factor equal to $\mu_0\sigma S$ ($S \equiv r^2$), $r$ is the distance between the source and the point of measurement, $\mu_0$ is the magnetic permeability of vacuum and $\sigma$ - electrical conductivity of the medium; s |
| $\tilde{x} = \frac{x}{\sqrt{S}}, \tilde{y} = \frac{y}{\sqrt{S}}, \tilde{t} = \frac{t}{t_0}$ | Non-dimensionless spatial coordinates $\tilde{x}, \tilde{y}$ and temporal variable $\tilde{t}$ |
| $I_0$ | Current strength, A |
| $t_{peak}$ | Time of arrival of pulse peak amplitude, s |
| $\tau_{co}$ | Pulse width of co-seismic unipolar pulses <br> Ref. Nenovski (2015) |
| $\tau_{pre}$ | Pulse width of pre-seismic unipolar pulses <br> Ref. Bleier et al (2009); Villante et al (2010) |
| $w_m$ | Apparent width (modelled and measured) of the magnetic field pulse, $B$ envelope |
| $w_{dm}$ | Apparent width (modelled and measured) of the magnetic field time derivative (d$B$/d$t$ envelope) |

**FIGURE CAPTIONS**

Fig. 1. Unipolar magnetic pulses recorded on March 18[th], 2009 both by fluxgate and overhauser magnetometers at the Aquila magnetic observatory. All pulses are negative and indicate a stable polarization property.

Fig. 2. Example of negative unipolar pulses observed by overhauser magnetometer at Aquila magnetic observatory on March 18[th], 2009. The pulse peak amplitude at 14:32 UT is of 1.2 nT, the next peak is ~0.7 nT.

Fig. 3. Co-seismic unipolar magnetic pulse observed at the 2009 Aquila earthquake main shock (after Nenovski, 2015). The signal is recorded by 1 Hz overhauser magnetometer.

Fig. 4. Unipolar pulses recorded at Tacna, Peru, 23 April, 2010 (after Dunson et al, 2011).

Fig. 5. Geometry of the electric current source. An element of the current volume of rectangular cross-section with sides $2x_o$ and $2y_o$ is sketched. Alongside a fault plane (strike and dip) is positioned. The arrow illustrates the assumed current direction (along +z). In general, the electric current source may have arbitrary volume $V$, cross-section area $S$ and orientation with respect to fault plane.

Fig. 6. Envelopes of the magnetic field pulse and its derivative assuming a strip cross-section current of impulse form. Amplitude and time axes are normalized.

Fig. 7. Envelopes of the magnetic field pulse and its derivative assuming a squared cross-section current of impulse form. Amplitude and time axes are normalized.

Fig. 8. Zone of applicability of the line concept and squared cross-section currents depending on the diffusion time. The dash line marks the limit of the line concept and transition to the squared cross-section current model.

Fig. 9. Experimental magnetic field pulse (dotted cure) and modelled magnetic field pulse $B$ for squared cross-section current impulse.

Fig. 10. Experimental magnetic field derivative (dotted line) and modelled difference d$B$ (solid line) for squared cross-section current impulse.

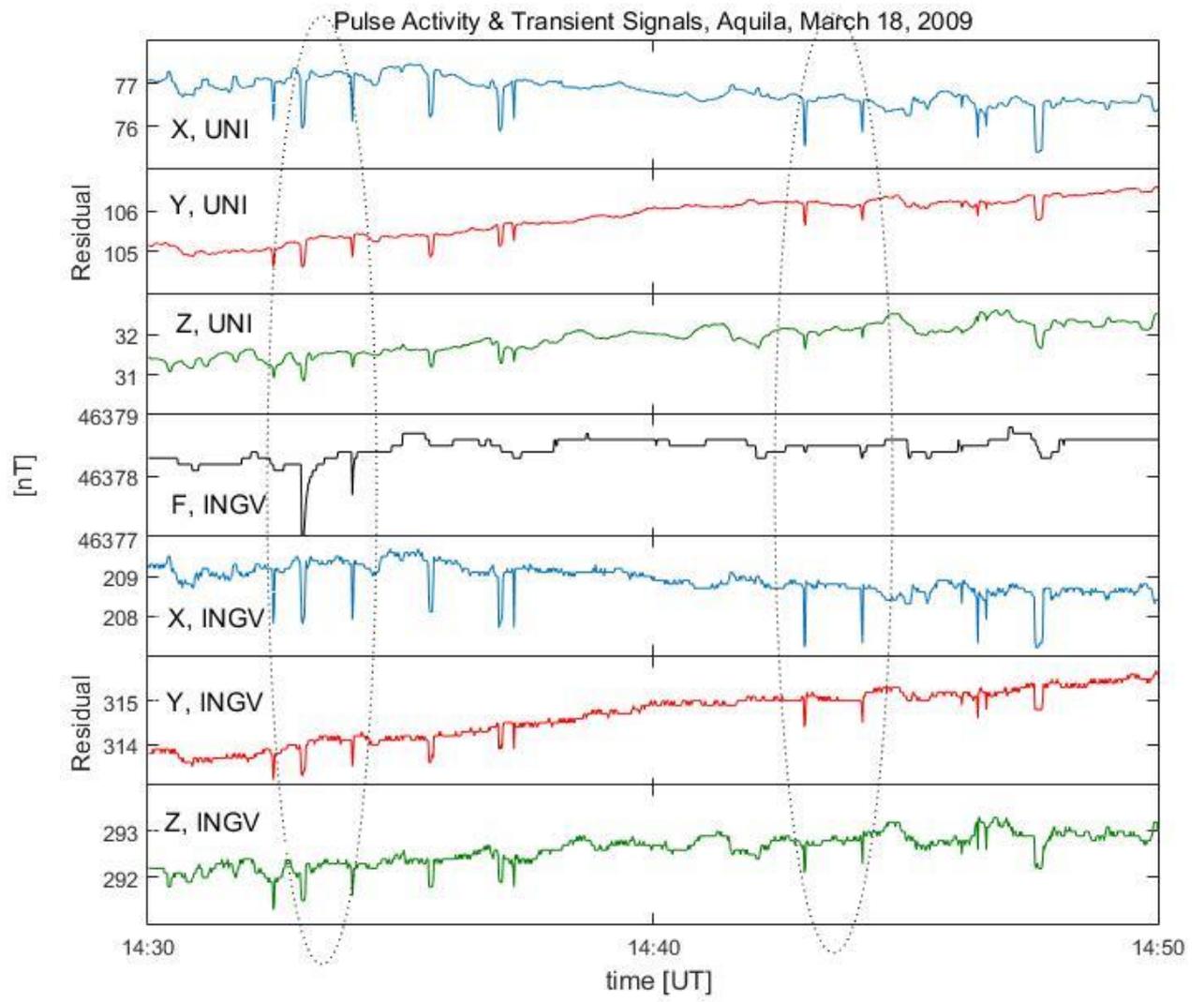

Fig. 1.

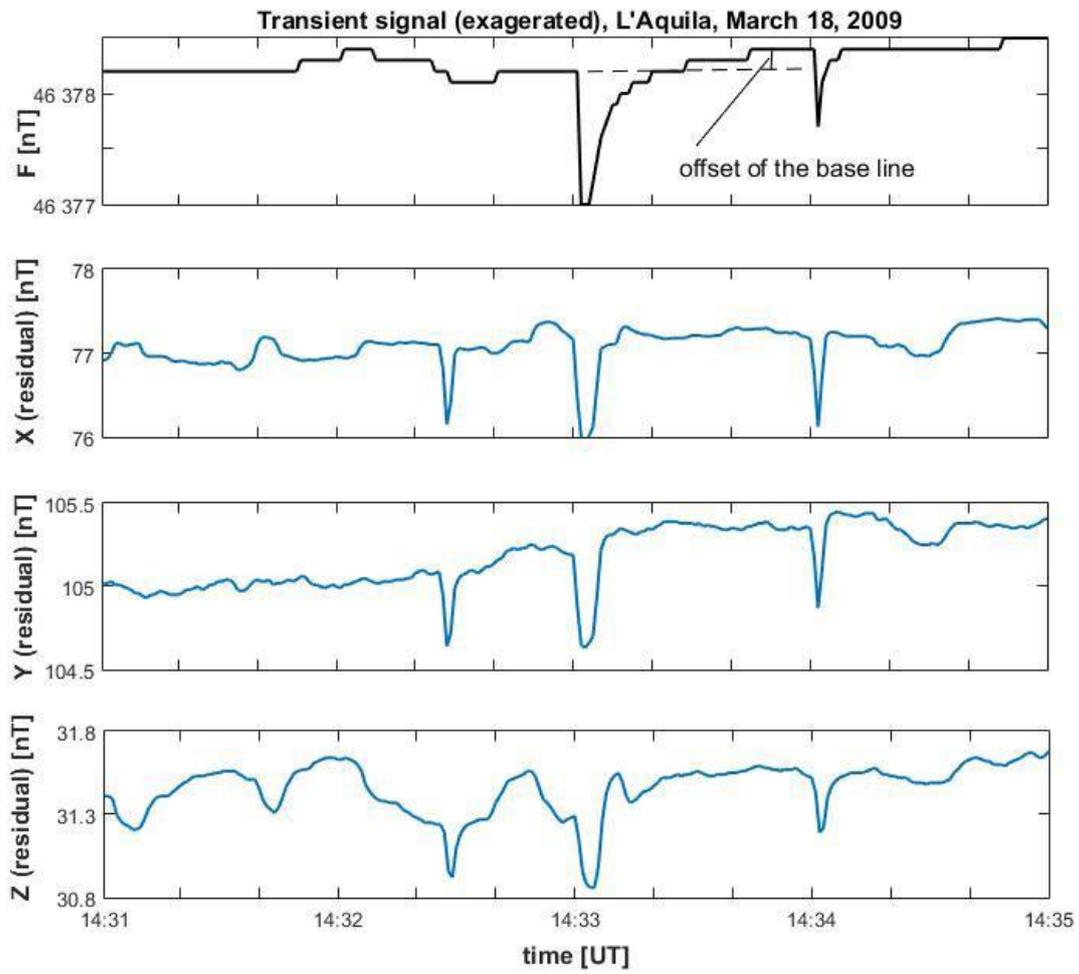

Fig. 2.

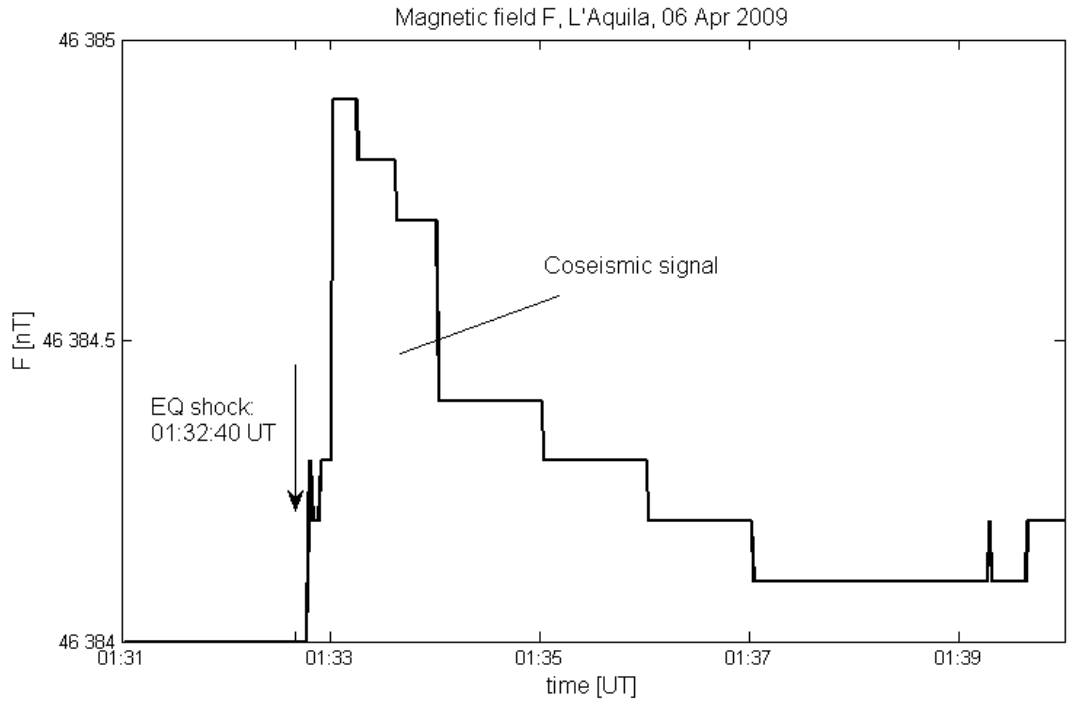

Fig. 3.

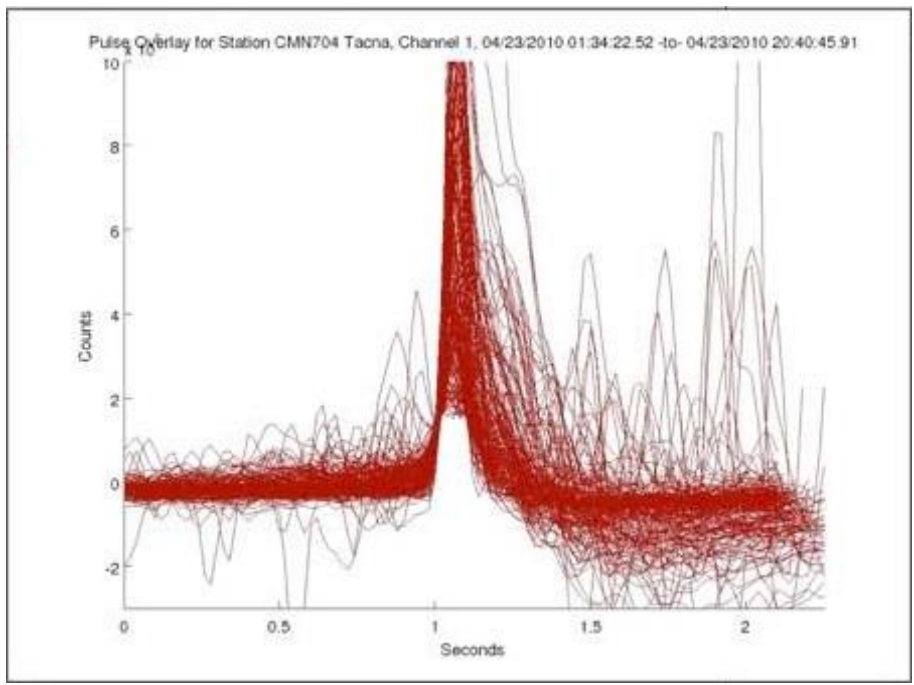

Fig. 4

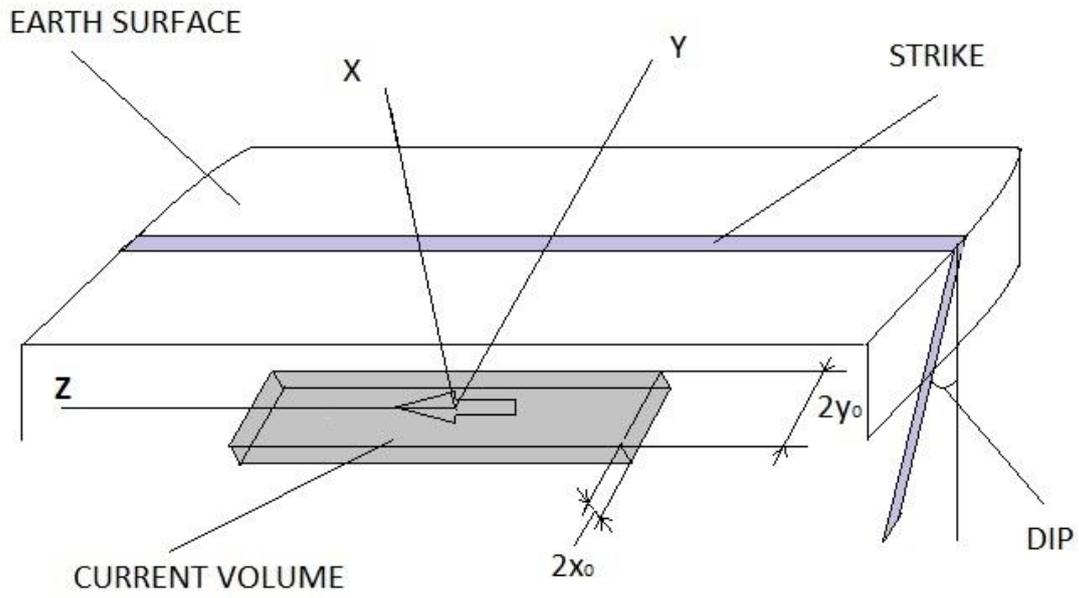

Fig. 5.

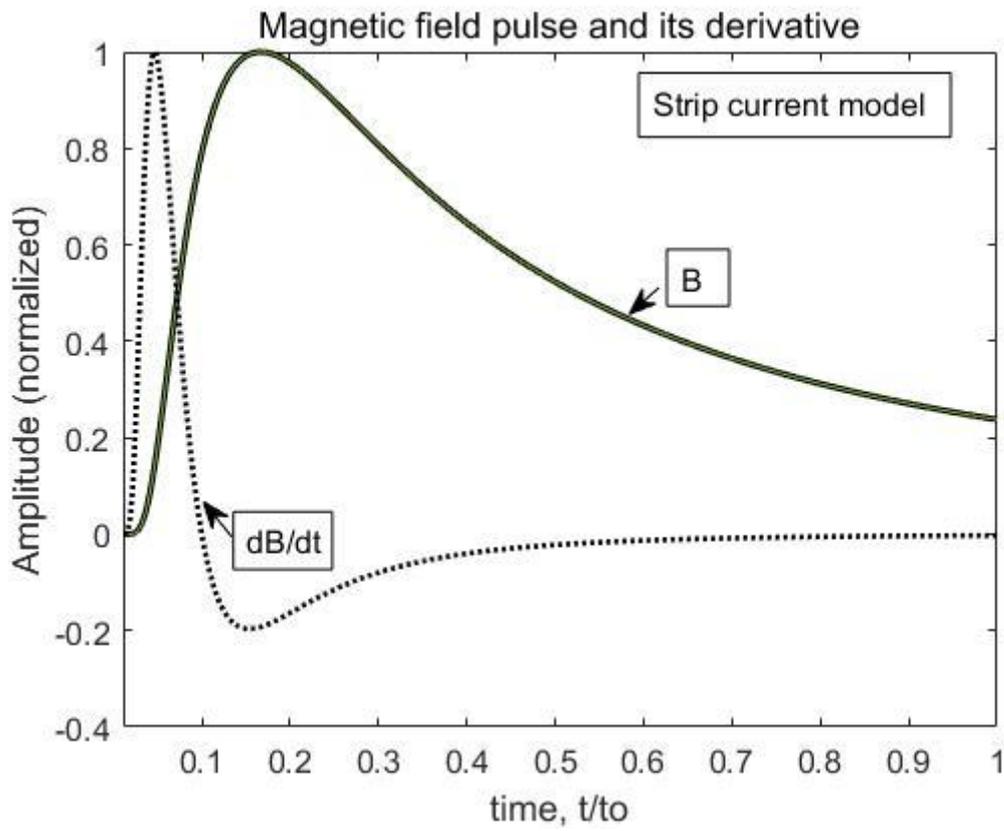

Fig. 6.

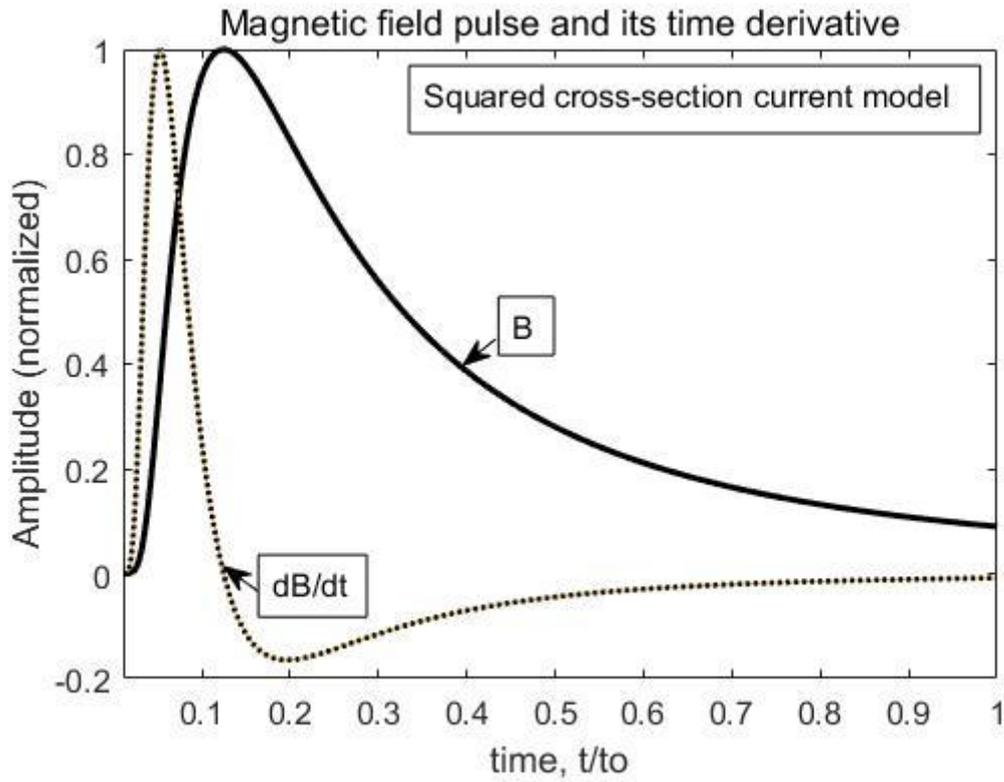

Fig. 7.

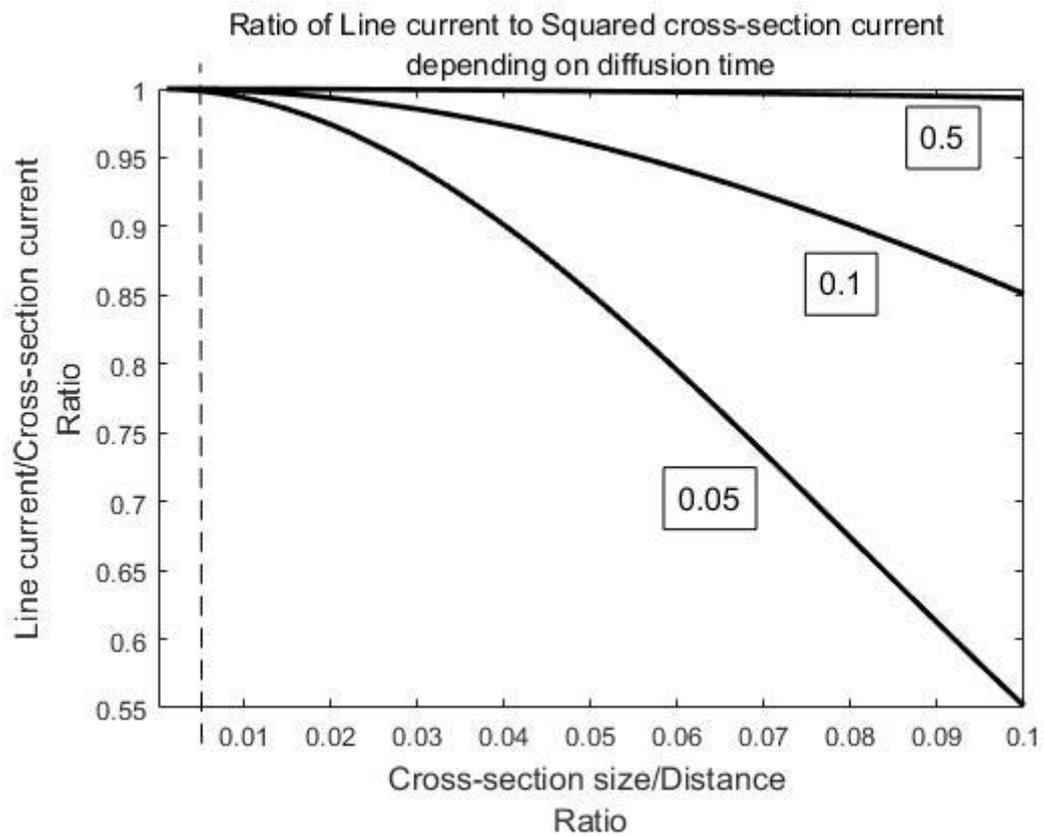

Fig. 8.

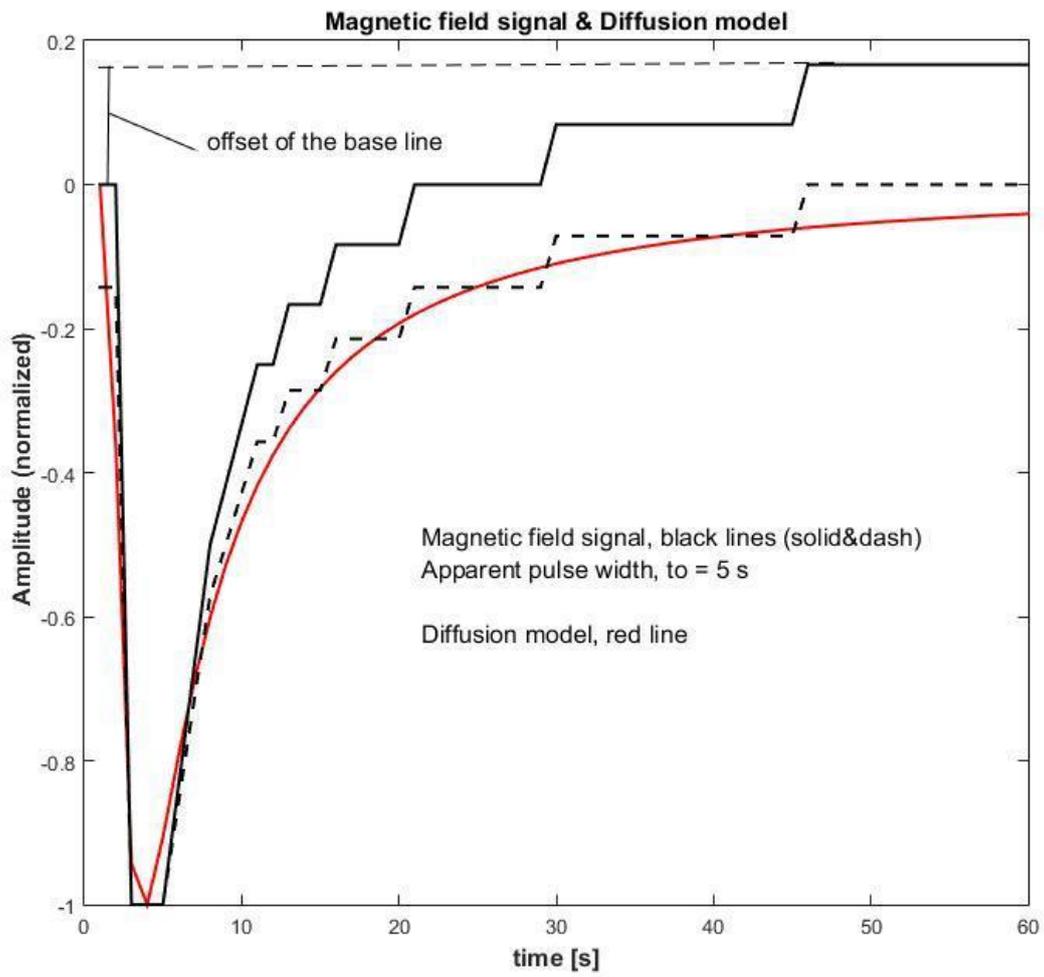

Fig. 9.

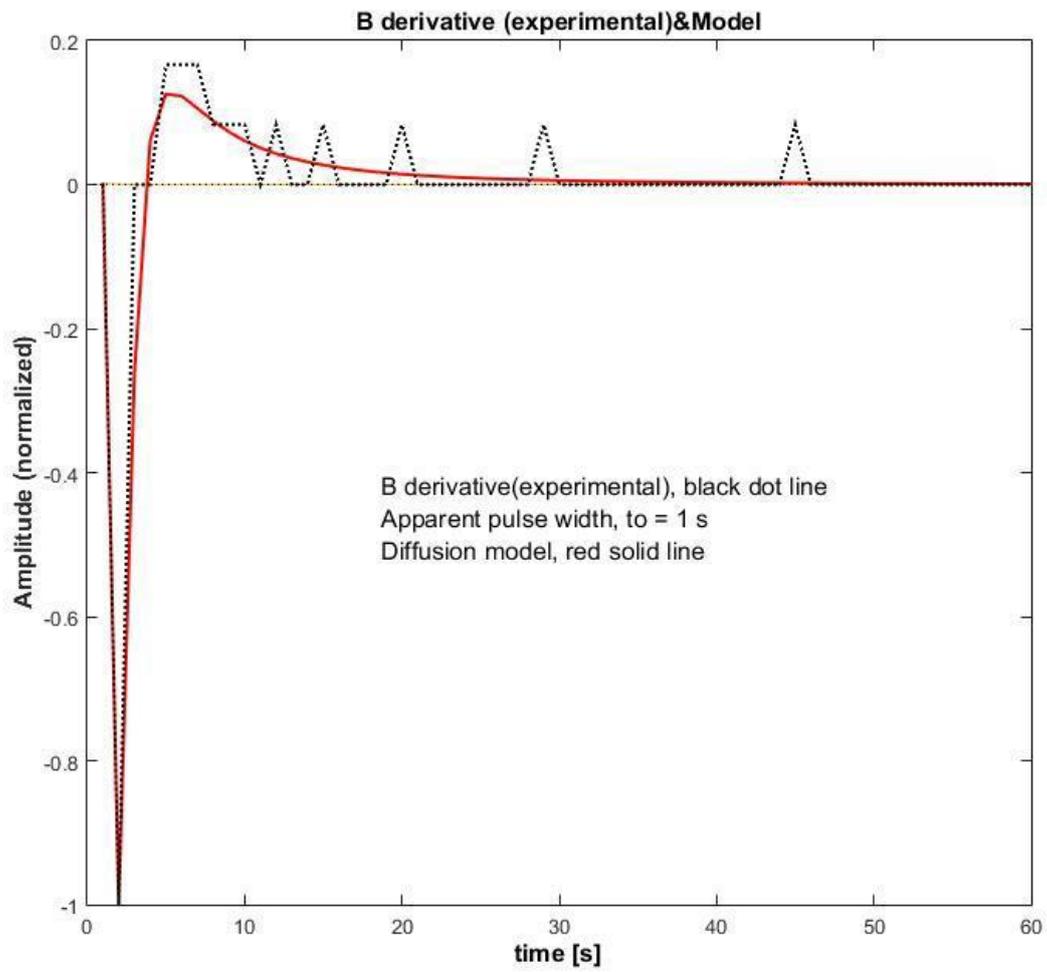

Fig. 10.